\documentstyle[aps,preprint]{revtex}

\def\bq{\begin{equation}}
\def\eq{\end{equation}}

\begin{document}

\draft

\title{Can a highly virtual nucleon experience
final state interactions in electron-nucleus scattering ?}

\author{Omar Benhar$^{\dagger}$ and Simonetta Liuti$^{* \dagger}$}

\address{$^{\dagger}$
INFN, Sezione Sanit\`a. Physics Laboratory, Istituto Superiore
di Sanit\`a.\\ Viale Regina Elena, 299. I-00161 Rome, Italy. \\
$^{*}$
Institute of Nuclear and Particle Physics, University of
Virginia.\\Charlottesville, Virginia 22901, USA.}

\date{\today}

\maketitle

\begin{abstract}

We discuss how the virtuality of the struck particle may affect
 the final state interactions in electron-nucleus scattering.
The extent to which short range correlations inhibit
rescattering taking place within the range of the repulsive
core of the
NN interaction is quantitatively analyzed.
The possible modifications of the nucleon-nucleon scattering
amplitude associated with the virtuality is also studied, within the framework
of a nonrelativistic model. The results suggest that the on shell
approximation can be safely employed in the kinematical region relevant to the
analysis of the available inclusive data at large momentum transfer and low
energy loss.

\end{abstract}

\pacs{PACS numbers: 25.30.Fj, 13.60.Hb, 24.10.Cn}

\narrowtext

Inclusive electron-nucleus scattering in the quasielastic regime, at
large momentum transfer q and low electron energy loss $\omega$
(corresponding to values of the Bjorken
scaling variable $x = Q^2/2m\omega > 1.5$), has
long been recognized
as a unique tool to measure the high momentum components of the nuclear wave
function or, equivalently, to get information on its behaviour at short
interparticle distance in configuration space \cite{cg}.
It is argued that, at large momentum transfer, electron-nucleus scattering
reduces to the incoherent sum of elementary processes in which the
electron scatters off individual nucleons, whose momentum ${\bf k}$ and
removal energy $E$ are distributed according to the spectral function
$P({\bf k},E)$, and the final state interactions (FSI) between the struck
particle and the spectator nucleons are negligibly small.
Within this picture, generally referred to as plane wave impulse
approximation (PWIA), the inclusive cross sections are expected
to exhibit a scaling behaviour in the variable $y$ \cite{3hescal,2hscal}.

The available data \cite{2hdata,3hedata,NE3,nmdata,NE3new}, extending up to
$Q^2 \sim 3\ (GeV/c)^2$ for
heavy targets, show sizeable $y$-scaling violations, indicating that the
PWIA regime has not been reached and FSI effects are still appreciable.
The failure of PWIA to describe the data has been also consistently confirmed
by
the existing
theoretical calculations of the inclusive cross sections, carried out using
realistic spectral functions
\cite{sauerpwia,gangofsix,cdl,bp,LDAnp}.
 Therefore, a quantitative understanding of FSI appears to be needed
 to explain the
measured cross sections and achieve a firm assessment of the longstanding
issue of high momentum components and short range correlations in nuclei.

A theoretical treatment of inclusive electron-nucleus scattering at large
momentum transfer including FSI effects, in which nucleon-nucleon ($NN$)
correlations are consistently taken into account in both the initial and the
final state within a microscopic many-body
theory, has been proposed in ref.\cite{gangofsix}. This
approach accounts for the available data on different
targets, ranging from deuteron to infinite nuclear matter\footnote{the
nuclear matter data have been obtained in
ref.\cite{nmdata}, fitting  the cross sections for finite $A$ to a mass formula
and  singling out the volume term.} \cite{gangofsix,bp,LDAnp}.
The main conclusion of refs.\cite{gangofsix,bp,LDAnp} is that most of the
strength observed in the low energy loss wing of the cross section, say at
$x>1.4$, comes from
processes in which the electron hits a nucleon  of low momentum  $k < k_F$,
$k_F$ being the Fermi momentum (in infinite  nuclear matter at equilibrium
density $k_F \sim .25\ GeV/c)$,  and the struck  particle undergoes FSI with
the $(A-1)$ spectator nucleons.

In a recent paper \cite{FS}, Frankfurt {\sl et al} have criticized the approach
of ref.\cite{gangofsix} claiming that, since it does not take into account
explicitly the virtuality of the struck particle, it is likely to overestimate
the effect of FSI.
If the struck nucleon is highly virtual, its lifetime, i.e. the time within
which it has to interact to be brought back on shell, may become comparable
with the range of the short range correlations ($\sim 1\ fm$), induced by the
repulsive core of the $NN$ force. On the other hand, $NN$ correlations make
the probability of finding a spectator to interact with at $t<1\ fm$ very
small.
 Hence, processes in which the struck particle is far off shell and undergoes
FSI are expected to be suppressed. Moreover, the authors of ref.\cite{FS}
argue that the amplitude describing the rescattering of the struck nucleon
may also be affected by its virtuality, leading to a further quenching of
FSI effects.

Whether a far off shell struck particle can undergo FSI
obviously depends upon the underlying nuclear dynamics, which dictates
the distribution in space of the spectators and the range of the rescattering
amplitude. In this paper we follow the approach of ref.\cite{gangofsix}, in
which
the dynamical quantity relevant to the description of FSI is the
eikonal propagator $U_q(t)$. The probability that the struck particle
undergo FSI during the time $t$ is given by $1-U^2_q(t)$.

In this paper, the structure of $U_q(t)$ is quantitatively analyzed, with the
aim of assessing the role of $NN$ correlations in suppressing FSI
of a far off shell, short lived, particle. We also estimate, within the
framework of a nonrelativistic model, the
difference between the imaginary parts of the nucleon self energy,
 obviously related to the imaginary parts of the $NN$
 scattering amplitude, evaluated on and off shell in the energy and momentum
range relevant to the analysis of the existing inclusive data.

In ref.\cite{gangofsix} the inclusive electron nucleus
cross section, i.e. the cross section of the process
$e + A\rightarrow e^\prime + X$, is written as a convolution integral:
\bq
\frac{d^2\sigma}{d\Omega d\omega}=\int d\omega^{\prime}
f_q(\omega-\omega^\prime)
\left(\frac{d^2\sigma}{d\Omega d\omega^\prime}\right)_{PWIA} ,
\label{folding}
\eq
where $(d^2\sigma/d\Omega d\omega)_{PWIA}$
denotes the PWIA cross section whereas the folding function $f_q(\omega)$,
embedding
all the information on FSI, is
the Fourier transform of the eikonal propagator $U_q(t)$, defined as
 (for the sake of simplicity we will refer to the case of infinite nuclear
 matter at equilibrium density $\rho$):
\bq U_q(t)=
exp \left\{ - \int_0^t d\tau~ \rho\int d^3r~g(r)~ w_q({\bf r} + {\bf v}\tau)
\right\}.
\label{U:def}
\eq
In eq.(\ref{U:def}), $g(r)$ is the nuclear matter radial distribution
function, while
${\bf v}$ denotes the velocity of the struck particle. The function
$w_q({\bf r})$ is related to the
imaginary part of the $NN$ scattering amplitude for incident momentum
${\bf p} = {\bf k} + {\bf q} \sim {\bf q}$,
which is known to be dominant
at large $q$:
\bq
w_q({\bf r})=\frac{2\pi}{m} \int \frac{d^3 p^\prime}{(2\pi)^3}~
e^{i{\bf p}^\prime{\bf r}}~Im f_q({\bf p^\prime})
\label{t:matrix}
\eq

The calculation of $U_q(t)$ according to eqs.(\ref{U:def})-(\ref{t:matrix})
requires two basic ingredients: i) $g(r)$, yielding
the probability that the struck particle hits one of spectators
after travelling a distance $r$, and
 ii) the amplitude $f_q({\bf p^\prime})$, describing the scattering process
 between the struck
nucleon and the spectators.

The results of ref.\cite{gangofsix,bp,LDAnp}, obtained using distribution
functions generated from realistic nuclear wave functions and the free
space amplitudes extracted from $NN$ scattering data \cite{NNampl}, show
that FSI is indeed significantly suppressed at small $t$ on account of
short range correlations. This feature is illustrated in fig.1, where the
eikonal propagator of a nucleon of momentum $p=1.95\ GeV/c$, evaluated with and
without
 inclusion of correlation effects (i.e. setting $g(r)=1$), are compared. It has
 to be kept in mind that the  tails
of the folding function $f_q(\omega)$ at $|\omega| > 0.3\ GeV$ are only
sensitive
to the shape of $U_q(t)$ at $t < 1 - 1.5\ fm$.

The calculations of refs.\cite{gangofsix,bp,LDAnp} also show that, in
 spite of the quenching produced by $NN$ correlations, the inclusion of
FSI produces a sharp rise of the cross section at $x>1.4$,  with respect to the
PWIA predictions, whose driving
mechanism  can be readily understood. In the
kinematical conditions under discussion, the PWIA nuclear response, defined
as:
\begin{eqnarray}
\nonumber
R({\bf q},\omega)  =  \int d^3k~dE~P({\bf k},E)~
 & \times &  \delta(\omega - E - \sqrt{|{\bf k}+{\bf q}|^2 + m^2} + m) \\
                  & \simeq & \frac{d^2\sigma(e+A \rightarrow e^\prime +
X)}{d^2\sigma(e+N \rightarrow e^\prime + X)}
\label{nucl:resp}
\end{eqnarray}
drops by three orders of magnitude as $x$ goes from 1 to 2. As
a  consequence, even a few percent deviation
from  unity of $U_q(t)$ at short $t$ ($t < 1 - 1.5\ fm$),
resulting in a tiny positive tail in $f_q(\omega)$ at large $|\omega|$
($ |\omega| \sim 0.3-0.5\ GeV$), dramatically affects the folded cross
section, moving strength from the quasi free bump ($x \sim 1$) to the
low $\omega$ region ($x > 1.5$) according to eq.(\ref{folding}).

The above argument can be rephrased in a somewhat more physical language.
The PWIA response in the region of the quasi free peak is dominated by
processes in which the electron hits a slow nucleon, whereas the strength
at $x> 1.5$ comes from scattering off nucleons of high momentum
($k >> k_F$), whose probability is strongly suppressed.
The results of ref.\cite{gangofsix,bp,LDAnp} essentially
show that the dominance of
low momentum nucleons in the nuclear ground state is so strong that, even
if the probability of FSI at short $t$ is quenched by $NN$
correlations, scattering off slow nucleons undergoing FSI
over a timescale of the order of $1\ fm$ is much more likely to occur
than scattering off fast nucleons. In fact, this turns out to be the leading
mechanism responsible for the inclusive cross section at $x > 1.5$, as
illustrated in fig.2. The PWIA nuclear matter response at
incident energy $3.6\ GeV$ and scattering angle $30^\circ$ (dash-dot line),
corresponding to $Q^2=2.3\ (GeV/c)^2$ at the quasi free peak $(x=1)$, is
shown as a function of $x$ and
compared to the results of the approach of  ref.\cite{gangofsix}. The solid
line corresponds to the full calculation, whereas the dashed line
has been obtained including in the calculation of FSI only the
contributions coming from nucleons of initial momentum
$k < k_F$, which appear to provide about $90\%$ of
the reponse  at $x > 1.4$.

To make clear that $R({\bf q},\omega)$
at $x>1.5$ is {\it only} sensitive
to FSI occurring within $\sim 1\ fm$, in fig. 2 we also show
the response evaluated from eq.(\ref{folding}) with a $U_q(t)$
which completely inhibits FSI at $t<r_c$, with $r_c \sim 1.1\ fm$. This
$U_q(t)$, shown by the diamonds
in fig.1, has been obtained using a steplike distribution function,
$g(r)=\theta(r-r_c)$, with $r_c$ fixed by the normalization, and a zero range
$NN$ amplitude in eqs.(\ref{U:def}) and (\ref{t:matrix}), respectively.
It appears that, at $x>1.5$, the diamonds come very close to the PWIA curve,
 showing that by artificially inhibiting the rescatterings occurring within
$\sim 1\ fm$ one kills the whole FSI effect.

Realizing that the inclusion of {\it realistic} $NN$ correlations does
not necessarily
results in a vanishing probability of FSI at $t < 1\ fm$ is very important
since, according to the estimates of ref.\cite{FS}, this
is the relevant timescale for FSI to occur in processes involving nucleons
 of low initial momentum at $Q^2$ of a few $GeV$ and low energy loss.
Assuming that $NN$ correlations totally inhibit FSI at $ t < 1\ fm$, would
therefore rule out scattering
off low momentum nucleons followed by FSI
as the relevant mechanism to move strength from the
quasi free peak to the large $x$ region. The results of
refs.\cite{gangofsix,bp,LDAnp} seem to indicate that this is not the case.

The question still remains, however, whether our calculation of the
eikonal propagator is strongly biased by the use of the on shell approximation
for the rescattering amplitude, i.e. whether the small $t$ behaviour
of $U_q(t)$ is strongly affected by
the differences between the free space scattering amplitude, employed
in the calculations, and the
amplitude describing the scattering of a far off shell nucleon.
This problem has been first raised in ref.\cite{UDS}.

Very little is known about the off shell behaviour of the $NN$ scattering
amplitude, or more generally of the nucleon self-energy, in the relativistic
regime relevant to the understanding of the inclusive data at high momentum
transfer. Therefore, one has to rely on the guidance provided by
nonrelativistic models.

The authors of ref.\cite{UDS} argue that the imaginary part of the self energy
of an off shell nucleon of momentum ${\bf p}$, related to
the $NN$ scattering $t$-matrix via
\bq
Im~ \Sigma(p,E) \sim \int \frac{d^3 p^\prime}{(2\pi)^3} n({\bf p}^\prime)~
\langle {\bf p}, {\bf p}^\prime |Im~ t(E) |
{\bf p}, {\bf p}^\prime \rangle ,
\eq
where $n({\bf p}^\prime)$ denotes the momentum distribution, can be sizeably
reduced with
respect to its on shell value $Im~\Sigma(p,p^2/2m)$. As an example, they
quote the result of ref.\cite{SM}, where the dependence upon $p$ and $E$ of the
self energy of a dilute hard sphere Fermi gas has been analyzed. The
 calculations of
ref.\cite{SM} show that for $E \sim E_F$, $E_F=k_F^2/2m$ being the Fermi
energy, and
$p>3k_F$, $Im~\Sigma(p,E)$ vanishes  due to energy and momentum
conservation. It has to be pointed out, however, that the
argument of ref.\cite{SM} does not apply to the kinematical domain
relevant to the study of FSI in electron nucleus scattering,
 corresponding to nucleon momenta  $p  \sim q \sim 2\ GeV/c$
 and energy $E \sim \omega_q/2$, with $\omega_q \sim q^2/2m$.

In order to get some insight in the behaviour of $Im~\Sigma(p,E)$ in the
relevant kinematical region, we have carried out a
calculation
following the approach of ref.\cite{BF}. $Im~\Sigma(p,E)$ has been
evaluated for infinite nuclear matter including
the second order Feynman diagram generally referred to as polarization
graph, which has been shown  to be responsible for most of the energy
dependence \cite{SM}, and using a Yukawa interaction. It has to be
stressed that
 this somewhat oversimplified treatment is expected to be adequate for the
purpose of the present study, since the energy
dependence of $Im~\Sigma(p,E)$ is
dominated by the phase space available for the decay of a quasiparticle
of momentum $p$ into a two-particle one-hole state.

In fig.3 we show the results of our calculation, in the form of the ratio
$Im~\Sigma(p,p^2/4m)/ Im ~\Sigma(p,p^2/2m)$, as a function of
the nucleon momentum $p$. It appears that the deviation from unity
rapidly decreases with $p$ for $p$ larger than $1\ GeV/c$, and
becomes less than $5\%$ at $p \sim 2\ GeV/c$, where the approach of
ref.\cite{gangofsix} is expected to be applicable.

A similar conclusion has been reached by Rinat and Tarragin in ref.\cite{RT},
where
 the off shell $t$-matrix for a nonrelativistic potential model has been found
to
approach the on shell one for momenta around $2\ GeV/c$.

The different results obtained in ref.\cite{UDS}, whose authors find
a large off shellness effect for momenta $p \sim 1 - 2\ GeV$ , have
probably to be ascribed to the particular
parametrization of $Im~\Sigma(p,E)$ employed in their calculations:
\bq
Im~\Sigma(p,E) = Im~\Sigma(\sqrt{2mE},E)~ e^{- \frac{b^2}{4}(p^2-2mE)}.
\label{sigma:diep}
\eq
Eq.(\ref{sigma:diep}) has been originally proposed in ref.\cite{MS} on the
basis of an
analysis of low energy nucleon-nucleus data. However, the authors of
ref.\cite{MS}
explicitly state that their prescription is not meaningful for
momenta much larger than $b^{-1}$. Hence, extending its use to
proton momenta in the $GeV/c$ region, using at the same time $b$ values
corresponding to $b^{-1}$ in the range $0.2 - 0.4\ GeV/c$, needed to get
a reasonable fit of the data, seems to be hardly justified.

In conclusion, the results of the present paper indicate that the treatment
of FSI proposed in ref.\cite{gangofsix}, in which correlation effects are
taken into account within a highly realistic many-body approach and the
rescattering process is described in the on shell approximation, is consistent
and applicable in the kinematical domain covered by the available inclusive
data at large $q$ and low $\omega$. The fact that in this regime a struck
nucleon of low initial momentum is sizeably off shell does not appear to
affect the main conclusion of refs.\cite{gangofsix,bp,LDAnp}, where the
strength at $x>1.4$ was ascribed mostly to electron scattering off slow
nucleons undergoing FSI.

Nonrelativistic many body calculations
suggest that the possibility that the eikonal propagator
be strongly affected by the on shell approximation for the rescattering
amplitude is very unlikely at $q>1.5-2\ GeV/c$.
Furthermore, it appears that at $Q^2 \sim 2-3\ (GeV/c)^2$
and $x \sim 2$, FSI occurring within the
range of $NN$ correlations, far from being totally inhibited, contribute about
$90\%$ of the response, while rescattering processes taking place over a
longer time scale do not play a significant role.

\acknowledgments
The authors are grateful to Prof. S. Fantoni for a number of discussions and a
critical reading of this paper. The hospitality of the Institute of Nuclear
and Particle Physics at the University of Virginia, where this work has been
completed, is also acknowledged.

\begin{figure}
\caption{Eikonal propagator of a nucleon of momentum $p=1.95\ GeV/c$
in infinite nuclear matter. The solid
line shows the results of the full calculation, whereas the dashed line has
been obtained disregarding the effect of short range $NN$ correlations. The
diamonds
show the results of the model calculation described in the text.}
\end{figure}

\begin{figure}
\caption{Nuclear matter response function at incident energy
$E=3.595\ GeV$
and scattering angle $\theta=30^\circ$, corresponding to $Q^2=2.3\ (GeV/c)^2$
at the quasifree peak. Dot-dash line: PWIA; solid line: full calculation
including the FSI; dashed line: FSI included only for nucleons
of initial momentum $k<k_F$; diamonds: same as the dashed line, but
with the FSI calculated using the $U_q(t)$
represented by the diamonds in fig.1.}
\end{figure}

\begin{figure}
\caption{Ratio $Im~\Sigma(p,p^2/4m)/Im~\Sigma(p,p^2/2m)$, in infinite
 nuclear matter at equilibrium density, shown as a function
of the nucleon momentum $p$.}
\end{figure}

\end{document}